\documentclass[preprint,showpacs,amsmath,amssymb]{revtex4}

\usepackage{dcolumn}
\usepackage{bm}

\begin{document}


\title{Angular Momentum of a Brane-world Model}

\author{JIA Bei$^{1,2}$,
 LEE Xi-Guo$^{1,3}$ and ZHANG Peng-Ming$^{1,3}$}
\affiliation{\footnotesize{$^1$ Institute of Modern Physics, Chinese Academy of Sciences, P.O.Box 31 Lanzhou, 730000, China\\
$^2$ Graduate University of Chinese Academy of Sciences, Beijing, 100080, China\\
$^3$ Center of Theoretical Nuclear Physics, National Laboratory of
Heavy Ion Collisions, P.O.Box 31 Lanzhou, 730000, China\\}}


\begin{abstract}
In this paper we discuss the properties of the general covariant
angular momentum of a five-dimensional brane-world model. Through
calculating the total angular momentum of this model, we are able to
analyze the properties of the total angular momentum in the
inflationary RS model. We show that the space-like components of the
total angular momentum of the inflationary RS model are all zero
while the others are non-zero, which agrees with the results from
ordinary RS model.
\end{abstract}



\maketitle

\section{Introduction}
Since the early works of Kaluza and Klein [1], the concept of extra
spacetime dimensions has been broadly consumed by physicists.
Recently the interests have been shifted from the traditional
Kaluza-Klein type to the so-called ``brane-world'' picture which is
inspired from the string theory. In brane models some fields (like
SM fields) are localized at a brane while other fields (such as
gravitation) can propagate in more dimensions. Numerous models of
this kind have been proposed for different purposes, such as large
extra dimensions models like the ADD scenario [2] and warped extra
dimensions models like Randall-Sundrem (RS) models [3]. These models
have different purposes ranging from solving the hierarchy problem
to symmetry breaking.

On the other hand, the understanding of different types of
conservation laws is very improtant in theories related to gravity,
such as the conservation laws of energy momentum and angular
momentum. In [4] Duan et al have proposed a generally covariant form
of the conservation law of energy-momentum using the orthonormal
frames, in which the energy-momentum is a covariant vector in
Riemannian spacetime. It is generally covariant and is able to
overcome the flaws in the expressions from Einstein and others. The
usage of this form of energy-momentum has been conducted within the
frame of RS model [5]. The result is generalized to more general
conditions [6], which reflects the gauge hierarchy problem from a
gravitational point of view. Following the similar procedure, Duan
and Feng proposed a covariant conservation law of angular momentum
[7], which is used to analyze the angular momentum conservation law
in the RS model [8].

In this paper we analyze the angular momentum of a general
brane-world model with one extra dimension. First in Section 2 we
discuss the genreal setup of the model and  some of the properties
of the angular momentum in the RS model which are from [7]. Then in
Section 3 we apply the method to calculate the angular momentum of
this general brane model and a specific example --- the inflationary
RS model, which is a generalization of the original RS model.
Finally in section 4 we present a conclusion of this paper.

\section{The Setup and the Angular Momentum Conservation}
We follow the procedure of the RS model that we have a
five-dimensional spacetime with two 3-branes in it. The fifth
dimension which is labeled as $y$ is compactified as $S^1/\
\mathbb{Z}_2$. The two 3-branes are located at the two fixed points
of the orbifold $y=0$ and $y=\pi$. From a cosmological point of
view, the 3-branes should be spatially homogeneous and isotropic.
Furthermore, we assume that the usual three-dimensional space is
also spatially flat. This gives us a general five-dimensional metric
of the form [9]

\begin{equation} \label{eps}
ds^2=-n^2(t,y)dt^2+a^2(t,y)\delta_{ij}dx^{i}dx^{j}+b^{2}(t,y)dy^{2}
\end{equation}\\
The total action of this system is then

\begin{equation} \label{eps}
S=\int d^4x dy \sqrt{-g}\ [2M^3R-\Lambda]+\sum_{i=1,2}\int d^4x
\sqrt{-g^{(i)}}[\mathcal {L}_i-\Lambda_i]
\end{equation}\\
where $g_{MN}$ and $R$ denote the five-dimensional metric and Ricci
scalar respectively, $\Lambda$ and $\Lambda_i$ are the cosmological
constants of that bulk and the branes, and $g^{(i)}_{\mu\nu}$ is the
induced metric on the branes. The above Latin letters of $M,N$ stand
for the five-dimensional indices. The signature of $g_{MN}$ is
$(-++++)$. We have separated the gravitational part and the matter
part of the action.

The construction of general relativity using the orthonormal frames,
i.e. the vielbeins, has been discussed by many authors. The local
frames represented by the vielbeins can be interpreted as a
beautiful tool of expressing the equivalence principle. In [7,10]
this method is used to analyze the covariant conservation law of
energy-momentum and angular momentum. For energy-momentum
conservation we need the general displacement transformations, while
for angular momentum the conservation may be obtained using the
local Lorentz invariance [8]. For our present model the form of the
background metric of the spacetime manifold and the total action is
a generalization of the original RS model, so from [8] it is shown
that there exists a conserved total angular momentum tensor density
with a superpotential

\begin{align} \label{eps}
&\partial_M (\sqrt{-g}j^M_{ab})=0\\
j^M_{ab}&=2M^3\partial_N (\sqrt{-g}V^{NM}_{ab})
\end{align}\\
Equation (3) is just the covariant conservation of the total angular
momentum density, while the superpotential can be expressed as

\begin{equation} \label{eps}
V^{MN}_{ab}=e^M_ae^N_b-e^M_be^N_a
\end{equation}\\
where $e^M_a$ is the vielbein. Therefore the superpotential and the
angular momentum density have the following properties

\begin{align} \label{eps}
V^{MN}_{ab}=&-V^{NM}_{ab}=-V^{MN}_{ba}\\
j^M_{ab}&=-j^M_{ba}
\end{align}\\
This total angular momentum density includes the spin density of the
matter fields. It can be shown that the total conservative angular
momentum can be obtained from

\begin{equation} \label{eps}
J_{ab}=\int_{\Sigma_t}
ej^M_{ab}d\Sigma_M=2M^3\int_{\partial\Sigma_t}e
V^{MN}_{ab}d\sigma_{MN}
\end{equation}\\
where $\Sigma_t$ is a cauchy surface of the spacetime manifold, $e$
is the determinant of the vielbein, and $ed\Sigma_M$ is the
covariant surface element of $\Sigma_t$ with
$d\Sigma_M=\frac{1}{4!}\ \epsilon_{MNOPQ}dx^N\wedge dx^O\wedge
dx^P\wedge dx^Q$ and $d\sigma_{MN}=\frac{1}{3!}\
\epsilon_{MNOPQ}dx^O\wedge dx^P\wedge dx^Q$. The second part of
Equation (8) is obtained from Gauss's law. On the cauchy surface
$\Sigma_t$ we have $dt=0$, therefore the total angular momentum can
be expressed as

\begin{equation} \label{eps}
J_{ab}=\int_{\Sigma_t} ej^t_{ab}dx^1dx^2dx^3dy
\end{equation}\\
We can see that this total angular momentum has the property
$J_{ab}=-J_{ba}$.

\section{The Angular Momentum of the brane model}
Now let's analyze the angular momentum of our brane model. We can
write the metric with orthonormal frames of this model as

\begin{equation}\label{eps}
ds^2=-\hat{\theta}^0\otimes\hat{\theta}^0
+\hat{\theta}^1\otimes\hat{\theta}^1
+\hat{\theta}^2\otimes\hat{\theta}^2
+\hat{\theta}^3\otimes\hat{\theta}^3
+\hat{\theta}^4\otimes\hat{\theta}^4
\end{equation}\\
so that the components of the orthonormal frames are

\begin{equation}\label{eps}
e^0_t=n(t,y),\             e^{i}_{x^i}=a(t,y),\ e^4_y=b(t,y)
\end{equation}\\
Then from Equation (5) we can get the non-zero components of
superpotential of our model

\begin{align}\label{eps}
&V^{tx^i}_{0i}=V^{x^it}_{i0}=-V^{x^it}_{0i}=-V^{tx^i}_{i0}=e^t_0e^{x^i}_i=\frac{1}{na}\nonumber\\
&V^{ty}_{04}=V^{yt}_{40}=-V^{yt}_{04}=-V^{ty}_{40}=e^t_0e^{y}_4=\frac{1}{nb}
\end{align}\\
With the superpotential we can calculate the total angular momentum
density. The non-vanashing components are

\begin{align}\label{eps}
&j^t_{04}=-j^t_{40}=-6M^3a^2a'\nonumber\\
&j^{x^i}_{0i}=-j^{x^i}_{i0}=-2M^3\partial_t(a^2b)\\
&j^{y}_{04}=-j^y_{40}=-6M^3a^2\dot{a}\nonumber
\end{align}\\
Finally we are able to obtain the two non-zero components of the
total angular momentum of this brane model which is

\begin{equation}\label{eps}
J_{04}=-J_{40}=6M_3\mathcal {V}\int nba^5a'dy
\end{equation}\\
where $\mathcal {V}$ represents the volume of the three-dimensional
space. We can see in this model all space-like components of the
total angular momentum are zero, which is a generalization of the
results in [8]. If we plug in

\begin{equation}\label{eps}
a=n=e^{-kr|y|},\qquad b=r
\end{equation}\\
then we recover the asymptotic behavior in [8].

Now let's focus on one specific model of our kind --- the
inflationary RS model [11,6]. The metric is

\begin{equation} \label{eps}
ds^2=\biggl(\frac{H_0A_0}{k}\biggr)^2\sinh^2
(-kB_0|y|+c)[-dt^2+e^{2H_0t}\delta_{ij}\ dx^{i}dx^{j}]+B_0dy^{2}
\end{equation}\\
which means that

\begin{equation}\label{eps}
a=\frac{A_0H_0}{k}\ e^{H_0t}\sinh (-kB_0|y|+c),\qquad
n=\frac{H_0}{k}\sinh (-kB_0|y|+c),\qquad b=B_0
\end{equation}\\
where $H_0=\frac{\dot{A}}{A}$, $k=\sqrt{\frac{-\Lambda}{24M^3}}$,
and $c$ is an integration constant which is related to other
parameters by

\begin{equation} \label{eps}
k_1=k\coth(c),\qquad -k_2=k\coth\biggl(-kB_0\pi+c\biggr)
\end{equation}\\
where $k_i=\Lambda_i/24M^3$. In [11] it is shown that the gauge
hierarchy is a general property of both the inflationary RS model
and the ordinary RS model, and in [6] the gauge hierarchy problem is
analyzed from a gravitational point of view using the orthonormal
frame method.

From Equation (16) we are able to get the non-zero components of the
total angular momentum of this inflationary RS model

\begin{equation}\label{eps}
J_{04}=-J_{40}=3M_3\mathcal {V}A^6_0(\frac{H_0}{k})^7\sinh(-kB_0\pi
 r)\sinh(-kB_0\pi r+2c)
\end{equation}\\
Therefore, in addition to the general property of our
five-dimensional brane model that all space-like components of the
total angular momentum are zero, the non-space-like components are
infinity, which is the result of gravity from the warped extra
dimension. We can see that this is a general property of both the
inflationary RS model and the ordinary RS model.

Let us now discuss some of the asymptotic behaviors of this result.
If we set $r\rightarrow 0$, which means there is no extra dimension,
we can see that Equation (19) becomes zero. This is obvious since in
this case there is no effect from the gravity of the warped extra
dimension. In [6] it is argued that if $c$ is near $kB_0\pi r$, we
can get an extremely large difference between these two energy
densities. Here we can see that under this circumstance the general
property of the total angular momentum we get here still holds.

\section{Conclusion}
In conclusion, we have discussed the properties of the general
covariant angular momentum of a five-dimensional brane-world model.
Through calculating the total angular momentum of this model, we are
able to analyze the properties of the total angular momentum in the
inflationary RS model, whose static limit is the original RS model.
We show that the space-like components of the total angular momentum
of the inflationary RS model are all zero while the non-space-like
components are infinity, which is the result of gravity from the
warped extra dimension and agrees with the results from the ordinary
RS model.

\begin{acknowledgments}
This work is supported by the CAS Knowledge Innovation Project
(No.KJCX3-SYW- N2,No.KJCX2-SW-N16) and the National Natural Science
Foundation of China (10435080, 10575123, 10604024).
\end{acknowledgments}


\end{document}